\documentclass[12pt,preprint]{aastex}

\shorttitle{}
\shortauthors{}
 
\begin{document}
 
\title{The mass spectrum of metal-free Stars 
resulting from photodissociation feedback: 
A scenario for the formation of low-mass population III stars}
\author{Kazuyuki Omukai \altaffilmark{1,2}
and Yuzuru Yoshii \altaffilmark{3,4}}
\altaffiltext{1}{Department of Physics, Denys Wilkinson Building, 
Keble Road, Oxford, OX1 3RH, UK}
\altaffiltext{2}{Division of Theoretical Astrophysics, 
National Astronomical Observatory, Mitaka, Tokyo 181-8588, Japan}
\altaffiltext{3}{Institute of Astronomy, Graduate School of Science, 
University of Tokyo, Mitaka, Tokyo 181-0015, Japan}
\altaffiltext{4}{Research Center for the Early Universe, Graduate School 
of Science, University of Tokyo, Bunkyo-ku, Tokyo 113-0033, Japan}
\email{omukai@astro.ox.ac.uk; yoshii@ioa.s.u-tokyo.ac.jp}

\begin{abstract}

The initial mass function (IMF) of metal-free stars that form in the initial
starburst of massive (virial temperatures $\ga 10^{4}$K) metal-free
protogalaxies is studied. In particular, we focus on the effect of
H$_2$ photodissociation by pre-existing stars on the fragmentation mass scale,
presumedly determined by the Jeans mass at the end of the initial free-fall
phase, i.e., at the so-called ``loitering phase,'' characterized by the local
temperature minimum. Photodissociation diminishes the Jeans mass at the
loitering phase, thereby reducing the fragmentation mass scale of primordial
clouds. Thus, in a given cloud, far ultraviolet (FUV) radiation from the first star, which is
supposedly very massive ($\sim 10^{3}M_{\sun}$), reduces the mass scale for
subsequent fragmentation. Through a series of similar processes the IMF for
metal-free stars is established. If FUV radiation exceeds a threshold level, the
star-forming clumps collapse solely through atomic cooling. Correspondingly,
the fragmentation scale drops discontinuously from a few time $10M_{\sun}$ to
sub-solar scales. In compact clouds ($\la 1.6$kpc for clouds of gas mass
$10^{8}M_{\sun}$), this level of radiation field is attained, and sub-solar mass
stars are formed even in a metal-free environment. Consequently, the IMF becomes
bi-modal, with peaks at a few tenths of $M_{\sun}$ and a few times 10$M_{\sun}$. The
high-mass portion of the IMF, $\xi_{\rm high}(m_{\ast})$, is found to be a very
steep function of the stellar mass $m_{\ast}$, $\xi_{\rm high}(m_{\ast}) \propto
m_{\ast}^{-5}$. Therefore, the typical mass scale of metal-free stars is
significantly smaller than that of the very first stars. In the Appendix we study
the thermal instability in collapsing primordial prestellar clumps, and discuss
why the thermal instability occuring during the three-body H$_2$ formation does
not appear to manifest itself in causing further fragmentation of such clumps.
\end{abstract}
 
 
\section{Introduction}

Recent theoretical studies claim that the first stars are very massive,
typically about a few times $10^{2} M_{\sun}$ to $10^3 M_{\sun}$, on the basis that (i)
the fragmetation mass scale owing to H$_{2}$ cooling is large, about
$10^{3}M_{\sun}$ (Bromm et al. 1999, 2002; Abel et al. 2000) and (ii) the
absence of dust grain formation allows most of the fragment mass to accrete onto a
protostar without being halted by stellar radiation feedback (Omukai \& Palla
2001, 2003).
 
The most iron-deficient star yet observed, HE~0107-5240 ([Fe/H]=--5.3,
$M_{\ast}=0.8M_{\sun}$; Christlieb et al. 2002), however, presents a puzzle in
this regard (see Beers 2003 for a concise summary of the pertinent issues).
Shigeyama, Tsujimoto, \& Yoshii (2003) suggest that HE~0107-5240 was most
probably born as a metal-free Population III star, and has been metal-enriched
by accretion of interstellar gas, rather than being born as a Population II
star. Whether or not this picture is correct, the very existence of HE~0107-5240
demonstrates that subsolar-mass star formation is possible even in metal-free or
extremely metal-poor environments. If HE~0107-5240 was born as a Population II
star, the fragmentation mass scale can clearly be reduced to $\la M_{\sun}$ as
long as a significant amount of metals has been depleted into dust grains
(Schneider et al. 2003). Even so, if a completely metal-free star is found in
the future, another solution will be required. This possibility has motivated us
to examine whether it is feasible to form subsolar-mass stars from primordial
gas.
 
While the first stars were probably very massive, it is important to consider
whether {\it all} metal-free stars must be so massive. Note that metal-free 
stars might not be exclusively associated with the very first stars, although
theoretical studies often make this assumption (e.g., Nakamura \& Umemura 2001;
Abel et al. 2002; Bromm et al. 2002). The key is that, depending on the nature
of Population III stars, the fragmentation process of primordial clouds can be
significantly altered owing to stellar feedback, such as the UV radiation field,
the kinetic effect of stellar winds and supernovae, etc. In fact, in the
presence of a strong far ultraviolet radiation field, metal-free star-forming
clumps can collapse solely by atomic cooling, with H$_2$ formation prohibited by
photodissociation (Nakada \& Yoneyama 1976; Hasegawa, Yoshii, \& Sabano 1981;
Omukai 2001). Then, the fragmentation mass scale can be reduced to sub-solar
scales. 

In this paper, we study the evolution of a pregalactic cloud, including its star
formation history and the growth of a UV radiation field within it. In
particular, we focus on conditions suitable for low-mass star formation, and
consider the initial mass function (IMF) of metal-free stars.
 
Small halos whose virial temperature is less than 8000K cool only by H$_2$
rotational/vibrational line emission. After the very first
episode of star formation, the entire mass of molecular gas is photodissociated
by a single or a few massive stars, and subsequent star formation activity is
quenched as long as the photodissociating star is present (Omukai \& Nishi 1999;
Ciardi, Ferrara, Governato, \& Jenkins 2000; Glover \& Brand 2001; Machacek,
Bryan, \& Abel 2001). In order for star formation to continue, the pregalactic
cloud must be massive enough, i.e., have virial temperature $>8000$K or
correspondingly, a total mass $\ga 10^{8}M_{\sun}$. In such clouds, which we
consider in this paper, metal-free star clusters will form eventually as a
result of succesive star formation processes.
 
The IMF of those metal-free star clusters is of particular importance to the
subsequent evolution of the universe, e.g., the metal enrichment, reionization,
etc. Consequently, the metal-free stellar IMF has been considered by a number of
previous authors. Among them, Yoshii \& Saio (1986) studied it in the framework
of Silk (1977)'s IMF theory, which assumes the opacity-limited fragmentation
scheme of Hoyle (1953), whereby the fragmentation is supposed to continue until
the clump becomes optically thick. However, according to recent numerical
simulations, the fragmentation mass scale may instead be determined when the
isothermality breaks, {\it before} the opacity becomes important, i.e., at the
epoch when the clump temperature begins to increase (Bromm et al. 2002).

By analysing the fragmentation process of filamentary primordial clumps,
Nakamura \& Umemura (2001) and Bromm et al.(2002) obtained similar estimates for
the fragmentation epoch of low-density filaments ($n_{\rm H} \la 10^{6}{\rm
cm^{-3}}$). Such filaments fragment with masses of a few times $100M_{\sun}$.
Higher density filaments ($n_{\rm H} \la 10^{12}{\rm cm^{-3}}$) fragment, with
masses on the order of $1M_{\sun}$, when the clump becomes optically thick to
H$_2$ lines (see also Uehara et al. 1996; Uehara \& Inutsuka 2000). On these
grounds, they claimed that the metal-free IMF becomes bi-modal, with peaks
around a few times $100M_{\sun}$ and $1M_{\sun}$, assuming the initial density
of filaments are distributed homogeniously up to $>10^{6}{\rm cm^{-3}}$.
However, it is not clear whether the initial densities extend to such high
values in realistic situations. Indeed, the 3-D numerical simulations by Abel et
al. (2000; 2002) and Bromm et al. (1999; 2002) have not succeeded in producing
such low-mass fragments. 
 
Above all, we emphasize that much of the recent work on the metal-free IMF does
not take proper account of feedback effects from already-formed stars. Although
various forms of feedback effects can be pointed out, we consider here only the
contribution of the FUV radiation field arising from pre-existing stars as a
first step toward solution of this problem.
 
This paper is organized as follows. In \S 2, the effect of a FUV radiation field
on the fragmentation mass scale of primordial gas is discussed. In \S 3, the
evolution of a pregalactic cloud, and the mass spectrum of formed stars are
discussed using the result obtained in \S 2. In \S 4, a number of important
parameters for pregalactic chemical evolution are presented. A brief summary
and discussion is then presented in \S 5.  In the Appendix, the possibility of
fragmentation due to thermal instability is re-examined.

\section{Fragmentation of Primordial Gas Clumps in a Far Ultraviolet Radiation Field}

In this section, we study the effect of FUV radiation on the fragmentation mass
scale of primordial gas. We assume that a pregalactic cloud has an inhomogeneous
density structure, consisting of individual ``clumps''. As these clumps
collapse, they eventually fragment into dense cores (``fragments'') of mass
$m_{\rm frag}$. We consider here the evolution of each clump, in particular its
fragmentation mass scale.
 
Although the detailed dynamical evolution of such clumps is rather complex, in
general, clumps more massive than several times the Jeans mass are thought to
first collapse in a disk-like manner, and subsequently fragment into filamentary
clumps (Miyama, Narita, \& Hayashi 1987a,b). Finally, with further collapse, the
filamentary clumps fragment again into dense cores (Inutsuka \& Miyama 1997).
All these processes proceed approximately on the free-fall timescale. The
density inhomogeneity is amplified by self-gravity in the course of the
collapse. The central region experiences run-away collapse, but leaves the outer
envelope of lower density unevolved. The size, namely the diameter in the
spherical and filamentary case or thickness in the disk case, of the central
region is about the Jeans length $\lambda_{\rm J}$, regardless of the shape of
the clumps.
 
The thermal and chemical evolution of primordial clumps irradiated with a FUV
radiation has been investigated by Omukai (2001). Here, we use the same model as
in Omukai (2001), with minor modifications so that the collapse timescale is
changed to the free-fall timescale for the filamentary configuration. Hence, we
adopt $t_{\rm col}=1/\sqrt{4G \rho}$ (i.e., Inutsuka \& Miyama 1997) rather than
the equivalent expression for the spherical case, $\sqrt{3 \pi/32G \rho}$, where
$\rho$ is the central density of clumps. This change has only a small effect on
our results.
 
The overview of the model is as follows. In this model, the thermal and chemical
evolution for the central region of the clumps is computed. Within the central
region, whose length scale is taken as $\lambda_{\rm J}/2$, physical quantities
are assumed to be homogeneous. The density increases at the free-fall rate:
\begin{equation} 
\frac{d \rho}{dt}=\frac{\rho}{t_{\rm col}}.
\end{equation}

\noindent The thermal evolution is followed by the energy equation:
\begin{equation}
\frac{dU_{\rm th}}{dt}=\frac{P}{\rho^{2}}\frac{d \rho}{dt}-{\cal L}_{\rm net},
\end{equation}

\noindent where $U_{\rm th}$ is the thermal enegy per unit mass, and 
$P$ is the thermal pressure. The net cooling rate, ${\cal L}_{\rm net}$, consists
of the contributions from the radiative cooling of atomic hydrogen lines, molecular
hydrogen lines, continuum radiation from primordial gas, and the cooling/heating
associated with chemical reactions. 
The optical depth is evaluated by the length scale of the central region 
$\lambda_{\rm J}/2$.
In the case that the clump is optically thick,
absorption and scattering effects are taken into account by the escape
probability method. 
The radiation field from outside the clump is also attenuated with the same 
optical depth.
The main effect of radiation field is photodissociation of molecular hydrogen, 
which is the major coolant in primordial gas clumps.
Non-equilibrium chemistry between ${\rm H, H_2, H^{+}, H^{-}}$, 
and ${\rm H_2^{+}}$ is solved. 
Helium is neglected since it is thermally inactive in the considered 
temperature range.
For more details, we refer the interested reader to Omukai (2001). 
 
The radiation field is assumed to be produced only by stars inside the 
pregalactic cloud, 
while the possible contributions either from the extragalactic background 
radiation or from active galactic nuclei are neglected.
As discussed \S 3 below, only very massive ($40-2000 M_{\sun}$) stars are
considered as sources of this UV radiation. 
Since the effective temperature of such massive stars increases only moderately 
with stellar mass, for simplicity, we assume that the radiation field inside 
the cloud is approximated by a diluted blackbody at a single temperature 
$T_{\rm rad}$ and take $T_{\rm rad}=10^{5}$K, which is a typical value of 
metal-free very massive stars (Ezer \& Cameron 1971).
Also, the ionizing photons are assumed to be trapped in the 
HII region formed around the massive stars, and not to contribute to the 
average radiation field in the protogalaxies. 
Then, using a parameter $W$, the mean intensity can be expressed as
\begin{equation}
J_{\nu}=W B_{\nu}(T_{\rm rad}),
\end{equation}

\noindent for $h \nu < 13.6$eV, and $J_{\nu}=0$ for $h \nu > 13.6$eV.
The parameter $W$ is called the dilution factor, hereafter.

Figure 1 shows the temperature evolution of collapsing primordial clumps in the
presence of FUV radiation, plotted as a function of the density. The dilution
factors, $W$, are as listed in this figure. The initial temperature is taken
arbitrarily to be 300K at $0.1 {\rm cm^{-3}}$, because the evolution at higher
density, where the fragmentation occurs (at H number density $n_{\rm H}\ga
10^{4}{\rm cm^{-3}}$), is not affected by this choice.

First, we describe the temperature evolution for primordial clumps without and
external radiation field (e.g., Saslow \& Zipoy 1967; Matsuda, Sato, \& Takeda
1969; Yoneyama 1972; Yoshii \& Sabano 1979; Carlberg 1981; Palla, Salpeter, \&
Stahler 1983; Omukai \& Nishi 1998). As seen in Figure 1 (the curve labelled
$W=0$), the temperature of the clump first rises adiabatically, then starts
decreasing abruptly by forming sufficient H$_2$ for cooling. The temperature
decrease stops at $n_{\rm H}\sim 10^{2-3}{\rm cm}^{-3}$, where rotational
levels of molecular hydrogen reach the local thermodynamic equilibrium. After
that, the temperature increases gradually. The actual temperature increase in
this phase is slower than shown in Figure 1, owing to the slower collapse rate
until the formation of nearly spherical objects by fragmentation. We call this
slow collapse epoch the ``loitering phase'' (e.g., Bromm et al. 2002).
 
The loitering phase clearly plays a role in the fragmentation of primordial
clumps, and has been studied by Bromm et al. (1999, 2002) and Abel et al.(2000,
2002) by way of 3-D hydrodynamical simulation, and by Nakamura \& Umemura (2001)
by 2-D hydrodynamics assuming that filamentary configurations are formed. 
According to these studies, the Jeans mass at the loitering phase is 
imprinted as a typical mass scale for fragmentation. 
Fragmentation occurs around this point ($n_{\rm H} \sim 10^{4-5}{\rm cm}^{-3}$), and the typical fragment mass scale is the Jeans mass then.
 
Next, let us see the effect of FUV irradiation on the thermal evolution.
Under the presence of a UV radiation field, the formation epoch of H$_2$ is 
delayed until higher densities and temperatures are reached (see Figure 1).
The density and temperature at the loitering phase also increase.
The net effect of this is to reduce the Jeans mass at the loitering phase.
In other words, {\it the stronger the FUV radiation, the smaller the 
fragmentation mass scale}. 
 
If the radiation intensity is even more increased, say to the level of $W>1
\times 10^{-5}$ in terms of the dilution factor, the photodissociation (for
$n_{\rm H}\la 10^{4} {\rm cm}^{-3}$) and collisional dissociation (for higher
densities) prohibit formation of sufficient H$_2$ for cooling during the entire
prestellar collapse phase. These clumps cool only through atomic radiation
processes, namely, H line cooling for $n_{\rm H} \la 10^{7}{\rm cm}^{-3}$ and
H$^{-}$ bound-free transitions for higher densities (see the evolutionary path
for $W=1.1 \times 10^{-5}$ in Figure 1). The evolutionary paths for the clumps
with $W>1.1 \times 10^{-5}$ are almost identical to one another except for the
shallow ``pit'' around $n_{\rm H} \sim 10^{4}{\rm cm}^{-3}$ for $W=1.1 \times
10^{-5}$, which dissappears in the presence of stronger radiation. An important
aspect of the collapse by atomic cooling is that the temperature continues to
decrease up to $n_{\rm H} \sim 10^{16}{\rm cm}^{-3}$. Consequently,
fragmentation occurs at very high density, and sub-solar mass scale fragments
are formed.
 
Next let us evaluate the fragmentation mass scale inferred from the physical
conditions at the loitering phase. We define the loitering point as the instance
of the temperature minimum around the loitering phase. Fragmentation does not
take place immediately at the loitering point. Instead, the clump contracts
more, by about 2 orders of magnitude in density (e.g., Nakamura \& Umemura
2001), owing to inertia before the fragmentation. Taking this effect into
consideration, we evaluate the density and temperature at fragmentation as
follows. After the loitering point, the clump is assumed to fragment by
contracting 2 orders of magnitude in density. During this period, the
temperature is assumed to be constant for molecular cooling clumps, while for
atomic cooling clumps the temperature rises (as in Figure 1) because the latter
become optically thick before fragmentation. Using the density and temperature
at the time of fragmentation, the fragment mass scale is estimated as follows.
The wavelength of maximum growing mode for fragmentation is $\lambda_{\rm mgr} =
2\pi R_{\rm core}$, where the core radius of the filament $R_{\rm
core}=\lambda_{\rm J}/2$. Thus, the typical mass scale for fragments is given by
$m_{\rm frag}=\rho \pi R_{\rm core}^{2} \lambda_{\rm mgr}$, where $\rho$
is the density at the fragmentation (e.g., Uehara et al. 1996).

In order to collapse and fragment, the mass of the clumps must be 
more massive than the maximum Jeans mass attained during the collapse.
Therefore, only sufficiently massive clumps can give a birth to dense 
cores and stars, eventually.
Probably, in the pregalactic clouds, some clumps satisfy this condition 
and some do not.
Note that, in our model, as long as the clumps are massive enough, 
the fragmentation mass is only a function of the radiation field 
since we have adopted the same thermal evolution given in Figure 1 
for all those clumps.

Fragmentation mass scales computed in this way are shown in Figure 2 as a
function of FUV radiation intensity. 
In the case of no radiation field ($W=0$), the fragmentation mass is 
$m_{\rm first} \simeq 2000M_{\sun}$, which is consistent with 
the numerical simulations (e.g., Bromm et al. 2002).
The fragmentation scale decreases gradually from the initial value 
$m_{\rm first}$ as the radiation intensity increases. 
When the radiation exceeds the threshold level, $W_{\rm
noH2} \simeq 1.1 \times 10^{-15}$, a transition from the H$_2$-cooling triggered
collapse to the atomic-cooling one occurs. Correspondingly, the fragmentation
scale drops discontinuously from the transiton mass, $m_{\rm tr} \simeq 40
M_{\sun}$, to sub-solar scales, $m_{\rm low} \simeq 0.3 M_{\sun}$.

For an insight into possible values for $W$,
let us consider here the radiation field at distance $r$ from 
a star of radius $R_{\ast}$ and of effective temperature $T_{\rm eff}$.
The mean intensity there is 
\begin{equation}
J_{\nu}= \frac{R_{\ast}^2}{4 r^2} B_{\nu}(T_{\rm eff}).
\end{equation}
The dilution factor is thus given by 
\begin{equation}
W = \frac{R_{\ast}^2}{4 r^2}.
\end{equation}
Using the relation $L= 4 \pi R_{\ast}^{2} \sigma T_{\rm eff}^{4}$,
and noting the luminosity of massive stars $L$ is close to $L_{\rm Edd}$, 
the dilution factor can be expressed as
\begin{equation}
W = 3 \times 10^{-15} \left(\frac{T_{\rm eff}}{10^{5} {\rm K}} \right)^{-4}
\left(\frac{m_{\ast}}{50M_{\sun}} \right) 
\left(\frac{r}{3{\rm pc}} \right)^{-2} \left(\frac{L}{L_{\rm Edd}} \right), 
\end{equation}
where $m_{\ast}$ is the mass of the star.
Then, if the massive stars ($\ga 50M_{\sun}$) are predominantly formed
and their mean separation is $\sim$ a few pc, the threshold level $W_{\rm
noH2} \simeq 1.1 \times 10^{-15}$ can be reached.
In \S 3, we discuss whether this level of radiation field 
is reached during the evolution of pregalactic clouds. 

In this paper we assume that the dense clumps formed by fragmentation around the
loitering phase do not produce sub-fragments. The possibility of
sub-fragmentation due to thermal instability has been clamed by some authors
(Sabano \& Yoshii 1977; Yoshii \& Sabano 1979; Silk 1983; Yoshii \& Saio 1986).
In the Appendix, we re-examine this question via a linear perturbation analysis
of the gravitationally collapsing primordial gas. Although the thermal
instability does occur in the active phase of the three-body H$_2$ formation
reaction, the instability remains weak and does not appear to cause further
fragmentation of the clumps. Therefore, we do not consider the sub-fragmentation
by the thermal instability in the following.

\section{Star Formation in Pregalactic Clouds \\
and Establishment of the Stellar Mass Spectrum}

In \S 2 we discussed the evolution of subclumps in a pregalactic cloud, and
obtained the fragmentation mass scale as a function of the FUV radiation
intensity. Here, let us consider the evolution of pregalactic clouds as a whole.
The history of the build-up of the stellar radiation field and the consequent IMF within
a pregalactic cloud is examined. As discussed in \S 2, sub-solar mass stars are
formed if the FUV radiation field exceeds the threshold level, which is $W_{\rm
noH_2} \simeq 10^{-15}$ in terms of the dilution factor. The condition of the
cloud when this transition actually occurs is also examined.

\subsection{A Model for Pregalactic Clouds}

Suppose that pregalactic clouds consist of two-phase medium, i.e.,  
diffuse gas and denser clumps.
The clumps collapse and form stars eventually, as described in \S 2.
Since the physical condition of the clumps, e.g., initial density, 
temperature, etc., varies from clumps to clumps, 
the star formation epoch in those clumps would differ each other. 
The star formation rate in clumps, local radiation field, 
and other local quantities, do depend on the local condition 
in the clumps.
Here for simplicity, we assume that the quantities averaged over 
the entire cloud, e.g., the averaged star formation rate,  
are controlled by averaged global quantities of the cloud,
e.g., the averaged gas density.
Specifically, we use a one-zone model of pregalactic 
clouds, where the physical quantities such as the gas density, 
$\rho_{\rm gas}$, stellar density, $\rho_{\ast}$, and
stellar radiation field, $u_{\rm rad}$, are taken to be averaged values 
(see e.g., Yoshii \& Saio 1985). 

Let the initial gas mass be $M_{\rm cl}$, and the length
scale be $l$. The initial gas density is given by 

\begin{equation}
\rho_{\rm gas}(0)=\frac{M_{\rm cl}}{l^3}.
\end{equation}

\noindent The length scale of the cloud $l$ is assumed to be constant in time.  
While the fragments themselves continue their own collapse, 
the parental pregalactic cloud is kept in dynamical equilibrium 
by the kinetic energy of the fragments.
 
The basic assumptions of this model are (i) star formation proceeds on the
free-fall timescale, and (ii) the stellar mass scale coincides with the
fragmentation scale, which is a function of the radiation intensity, using the
relation in Figure \ref{fig:f2}.

The average gas density, $\rho_{\rm gas}$, in the cloud is followed by

\begin{equation}
\frac{d \rho_{\rm gas}}{dt}=-\frac{\rho_{\rm gas}}{t_{\rm sf}}.
\label{eq:dev}
\end{equation}

\noindent For the star formation timescale, $t_{\rm sf}$, we take the free-fall 
timescale, $t_{\rm col}=1/\sqrt{4 G \rho_{\rm gas}}$. We neglect the recycling
of gas ejected by either supernovae or stellar winds, assuming that it directly
escapes from the cloud. Also, gas is always assumed to be primordial and
chemical enrichment effecte are not taken into account, since we are now
interested only in metal-free stars.
  
Let the number of stars per unit volume at time $t$
in mass range $dm_{\ast}$ be $\xi(m_{\ast},t)dm_{\ast}$.
By summing up the radiation from all stars, the average radiation 
enegy density in the cloud is given by 

\begin{equation}
u_{\rm rad}=\frac{l}{c} \int \xi(m_{\ast},t)L(m_{\ast})dm_{\ast}.
\end{equation}

\noindent 
The values of luminosity $L(m_{\ast})$ of a Population III star 
with mass $m_{\ast}$ are taken from Schaerer (2002). 
Here, the radiation spectrum is supposed to be a diluted blackbody of 
$T_{\rm rad}=10^{5}$K, since we consider the radiation from massive metal-free
stars ($>m_{\rm tr} \simeq 40M_{\sun}$). The mass scale of fragmentation at time
$t$, $m_{\rm frag}(t)$, is determined by $u_{\rm rad}(t)$ by the relation shown
in Figure 2. Stars are assumed to turn off after their main sequence lifetimes.
Here, the main sequence lifetime of stars without stellar wind given by Schaerer
(2002) is used. For stellar masses higher than $500M_{\sun}$, whose lifetime is
not given in the literature, the same lifetime as that for $500M_{\sun}$ is
used.
 
With the above assumptions, the equation for stellar number density, 
$\xi(m_{\ast},t)$, is formally written as

\begin{equation}
\frac{ \partial \xi(m_{\ast},t)}{\partial t}=
\frac{\rho_{\rm gas}}{m_{\rm frag} t_{\rm sf}} 
\delta (m_{\ast}-m_{\rm frag})
-\xi(m_{\rm d},t_{\rm d}) \left| \frac{dm_{\rm d}}{dt} \right| 
\delta (m_{\ast}-m_{\rm d}),
\label{eq:nev}
\end{equation}

\noindent where $m_{\rm d}$ is the mass of stars that dies at time $t$, and $t_{\rm d}$
is their formation time, hence $t_{\rm d}=t-t_{\ast}(m_{\rm d})$.
In the equation above, the first term on the right hand side is 
the star formation rate, and the second term is the stellar death rate.
Equations (\ref{eq:dev}) and (\ref{eq:nev}) are integrated until 
$\rho_{\rm gas}$ falls below 10 \% of the initial value, $\rho_{\rm gas}(0)$,
or the FUV radiation exceeds the threshold value for atomic-cooling collapse,  
for which we use the threshold dilution factor 
$W_{\rm noH_{2}}=1 \times 10^{-15}$.
Note again that this is still a qualitative, rather than a quantitative, 
approach to study early star formation in pregalactic clouds, 
and it is of course not realistic that only 10 \% 
of the initial gas remains into the ISM.
The evolution of pregalactic clouds is not affected 
by this value: only the termination epoch of the calculation 
is changed.

\subsection{The Conditions for Low-Mass Star Formation}

Two cases of cloud mass $M_{\rm cl}$ are studied: (a) $M_{\rm
cl}=10^{7}M_{\sun}$, and (b) $10^{8}M_{\sun}$. For case (a), the length scale,
$l$, is varied from 100 to 1000pc, while for case (b), $l=300-3000$ pc. The
evolution of the average radiation field,$u_{\rm rad}$, is presented in Figures 3
(for case a) and 4 (for case b).
 
We see in Figures 3 and 4 that the radiation field in relatively compact clouds,
whose length scale $l \leq 500$pc for $M_{\rm cl}=10^{7} M_{\sun}$ or 1000pc for
$M_{\rm cl}=10^{8} M_{\sun}$, reaches the critical level for low-mass star
formation, while it does not in more diffuse ones. In pregalactic clouds where
low-mass stars are formed, the value of $W$ is expected to remain at the
threshold level $W_{\rm noH_2}$ until early-formed massive stars begin to die
about $t \simeq 2 \times 10^{6}$yrs, since low-mass stars do not emit FUV
radiation. On the other hand, in larger clouds the radiation enegy density
$u_{\rm rad}$ fails to reach the critical level $u_{\rm noH_2}=W_{\rm noH_2} a
T_{\rm rad}^{4}$, and declines after the first stars begin to turn off. In other
words, there is a threshold cloud length, $l_{\rm th}$, below which the radiation
intensity reaches the critical value. This threshold length is
larger for the cloud of $10^{8}M_{\sun}$ ($l_{\rm th} > 1000$pc) than for that
of $10^{7}M_{\sun}$ ($l_{\rm th} < 1000$pc).
 
We now discuss analytically the behavior of the radiation field and the threshold cloud
length. For simplicity, we suppose that the gas depletion timescale, $t_{\rm dep}
\simeq t_{\rm sf}(0)$, is longer than the lifetime of the first star,
$t_{\ast}(m_{\rm first})$. A similar discussion holds in the opposite case with
some modification. Also, since the stellar lifetime and luminosity-to-mass ratio
is only weakly dependent on mass for massive stars, in the following analytical
discussion we use a constant value of $t_{\ast, high}=2 \times 10^{6}$ yr and
$(L/M)_{\ast}=3 \times 10^{4} (L/M) _{\sun}$, which are the values for a star of
$1000M_{\sun}$, for all massive stars ($>m_{\rm tr}$).

Before significant gas depletion takes place, at $t<t_{\rm dep}$, the average gas density
$\rho_{\rm gas} \simeq\rho_{\rm gas}(0)$ and the star formation rate is
approximately constant at $\rho_{\rm gas}(0)/t_{\rm sf}(0)$. The stellar mass
density is 

\begin{equation}
\rho_{\ast}(t) \simeq 
\left\{
\begin{array}{ll}
\rho_{\rm gas}(0) t/t_{\rm sf}(0)
~~~~~t<t_{\ast, \rm high}, \\
\rho_{\rm gas}(0)  t_{\ast, \rm high}/t_{\rm sf}(0)
~~~~~t_{\ast, \rm high}< t < t_{\rm dep}. 
\end{array}\right.
\end{equation}

\noindent After the gas is depleted significantly, the star formation rate
drops to $\rho_{\rm gas}(t)/t_{\rm sf}(t)$.
Thus

\begin{equation}
\rho_{\ast}(t) \simeq 
\rho_{\rm gas}(t) t_{\ast, \rm high}/t_{\rm sf}(t) 
~~~~~t> t_{\rm dep}.
\end{equation}

\noindent By the way, the radiation density is given by

\begin{equation}
u_{\rm rad}(t)=\frac{l}{c} \rho_{\ast}(t) (L/m)_{\ast},
\end{equation}

\noindent as long as all the stars are massive, namely $u_{\rm rad}< u_{\rm noH2}$.
Since $(L/m)_{\ast}$ is only weakly dependent on $m_{\ast}$, the radiation field
is proportional to the stellar mass density $\rho_{\ast}$. Therefore, the
radiation intensity first increases linearly in time ($t<t_{\ast, \rm high}$),
then remains constant ($t_{\ast, \rm high}< t < t_{\rm dep}$), and finally
declines as the gas in the cloud depletes ($t> t_{\rm dep}$).

The radiation dilution factor $W$, which is related to the radiation 
energy density $u_{\rm rad}$ by $W=u_{\rm rad}/7.56 \times 10^{5}
{\rm (erg/cm^{3})}$, 
is

\begin{equation}
W=0.5 \times 10^{-14} \left( \frac{l}{1{\rm kpc}} \right)^{-7/2}
\left( \frac{M_{\rm cl}}{10^{8} M_{\sun}} \right)^{3/2}
\left( \frac{t}{2 \times 10^{6} {\rm yr}} \right)
~~~(t<t_{\ast, \rm high}), 
\end{equation}

\noindent and reaches maximum at $t=t_{\ast, \rm high}$.
Therefore, only in clouds whose length scale is smaller than the threshold
length

\begin{equation}
l_{\rm th}=1.6{\rm kpc} (M_{\rm cl}/10^{8}M_{\sun})^{3/7},
\label{eq:lth}
\end{equation}

\noindent the dilution factor $W$ reaches $W_{\rm noH_{2}} \simeq 1 \times 10^{-15}$, and
sub-solar mass stars are formed thereafter. This value of $l_{\rm th}$ agrees
well with the numerical result presented in Figures 3 and 4.
 
The number of low-mass stars formed can be estimated by the following
considerations. Since low-mass stars do not emit UV radiation, $u_{\rm rad}$
remains constant after reaching $u_{\rm noH_2}$ at $t=t_{\rm lmsf}$, where 

\begin{equation} 
t_{\rm lmsf}=4 \times 10^{5}{\rm yr} \left( \frac{l}{\rm 1kpc} \right)^{7/2} 
\left( \frac{M_{\rm cl}}{10^{8} M_{\sun}} \right)^{-3/2},
\end{equation}

\noindent until the massive stars start to die at $t=t_{\ast, \rm high}$.
Then, the duration of low-mass star formation is 
$t_{\ast, \rm high}-t_{\rm lmsf}$.
Multiplying by the star formation rate, the mass density of low-mass stars is  

\begin{equation}
\rho_{\ast,{\rm low}}= \frac{\rho_{\rm gas}(0)}{t_{\rm sf}(0)}
(t_{\ast, \rm high}-t_{\rm lmsf}).
\end{equation}

\noindent Metal-free star formation continues for the duration 
$\Delta t= t_{\ast, \rm high}$, i.e., from the first star formation until the
massive stars begin to explode as supernovae. At time $t=t_{\rm lmsf}$ in this
period, low-mass star formation starts. Since the star formation rate is
approximately constant at $\rho_{\rm gas}(0)/t_{\rm sf}(0)$, the mass fraction
of metal-free stars that is locked into low-mass stars is given by the ratio of
the two durations:

\begin{eqnarray}
f_{\rm low} 
&=&\frac{\rho_{\ast,{\rm low}}}{\rho_{\ast,Z=0}}
=\frac{t_{\ast, \rm high}-t_{\rm lmsf}}{t_{\ast, \rm high}} \\
&=&1- \left( \frac{l}{l_{\rm th}} \right)^{7/2} \\
&=&1-0.2 \left( \frac{l}{\rm 1kpc} \right)^{7/2} 
\left( \frac{M_{\rm cl}}{10^{8} M_{\sun}} \right)^{-3/2},
\label{eq:flow}
\end{eqnarray}

\noindent where $\rho_{\ast, Z=0}$ is the density of metal-free stars.

\subsection{The Stellar Initial Mass Function}

Because of the mass gap between low-mass stars ($\simeq 0.3M_{\sun}$) and 
high-mass stars ($\ga 40M_{\sun}$), the metal-free IMF becomes bi-modal. 
 
In our idealized model, the low-mass IMF is delta-function-like: only stars of
$m_{\rm low} \simeq 0.3M_{\sun}$ are formed (see Figure 2). Actually,
a distribution of the fragmentation mass around this value would be present.
Competitive accretion of ambient gas would create some distribution of the
protostellar mass. Also, the coagulation between fragments might form more
massive cores. As a result of these processes, the low-mass IMF would be formed,
although we do not discuss it further here. Recall that the peak ratio between
the high and low-mass IMF can be obtained by the mass fraction of low-mass stars
$f_{\rm low}$ given by equation (\ref{eq:flow}).
 
On the other hand, the high-mass IMF, $\xi_{\rm high}(m_{\ast})$, of metal-free
stars that are more massive than the transition mass $m_{\rm tr} \simeq
40M_{\sun}$ has been obtained in our model, and is shown in Figure 5 for clouds
of $M_{\rm cl}=10^{8}{M_{\sun}}$ (case a). In three cases of $l=300$, 500, and
1000pc, the shape of the IMF is identical, and the value of IMF at the same mass is
proportional to $l^{-1}$. In the case of $l=$3 kpc, the radiation field does not
reach the threshold value, and no low-mass stars are formed (see Figure 4). In this
case, the instantaneous mass of forming stars first reaches a minimum of
$140M_{\sun}$ and then increases as the radiation intensity declines. The stars
of $m_{\ast} \la 300M_{\sun}$ can be produced at two different epochs, 
first when the radiation density is increasing and second when the radiation
density is decreasing. The stars formed at different epochs are shown separately in Figure 5. The actual IMF is the sum of them.
 
We now present an analytic derivation of the high-mass IMF, $\xi_{\rm high}$,
following Silk (1977). Here, we consider stars formed at $t<t_{\ast}(m_{\rm
first})$, namely, during the time when the first star is still alive, as those formed later
are probably metal-polluted. Thus, the maximum mass of existing stars is just the
mass of the first star, $m_{\rm first}$. The radiation energy density at time $t$
is given by  

\begin{equation}
\frac{u_{\rm rad}(t)}{l/c}=\int^{m_{\rm first}}_{m_{\ast}(t)} 
\xi_{\rm high}(m_{\ast}')L(m_{\ast}')dm_{\ast}',
\label{eq:uimf}
\end{equation}

\noindent where $m_{\ast}(t)$ is the mass of forming star at $t$. 
Differentiating both sides of the equation (\ref{eq:uimf}) with $m(t)$, we
obtain the mass spectrum

\begin{equation}
\xi_{\rm high}(m_{\ast})=\frac{c}{l L(m_{\ast})} \left| 
\frac{du_{\rm rad}}{dm_{\ast}} \right|.
\label{eq:aimf}
\end{equation}

\noindent Suppose now the relations $u_{\rm rad} \propto m_{\ast}^{\alpha}$
and $L \propto m_{\ast}^{\beta}$. In this case, the IMF becomes

\begin{equation}
\xi_{\rm high} \propto m_{\ast}^{\alpha-\beta-1}.
\label{eq:bimf}
\end{equation}

\noindent This relation was first derived by Silk (1977).
From Figure 2, the approximate relation between fragmentation mass and radiation
density, $m_{\rm frag} \propto W^{-1/3}$, holds for massive stars of $m_{\ast}
\sim 100M_{\sun}$. For these massive stars, the luminosity-mass relation $L
\propto m_{\ast}$ holds. Then, it follows that $\alpha =-3$ and $\beta =1$ for
those stars. Substituting these values into relation (\ref{eq:bimf}), we
finally obtain the metal-free IMF for massive stars:

\begin{equation}
\xi_{\rm high} \propto m_{\ast}^{-5}.
\label{eq:cimf}
\end{equation} 

\noindent The mass spectrum becomes a very steep function of mass 
because of the negative value of $\alpha$. 
The normalization constant for the relation (\ref{eq:cimf}) depends 
on the length scale of the pregalactic cloud as $\propto l^{-1}$, 
as seen in equation (\ref{eq:uimf}).
Thus, using the values in Figure 5, the IMF can be expressed apprximately as 
\begin{equation}
\xi_{\rm high} = 3 \times 10^{-6} ({\rm stars}~M_{\sun}^{-1}~{\rm pc}^{-3})
\left(\frac{l}{1{\rm kpc}}\right)^{-1} 
\left(\frac{m_{\ast}}{50M_{\sun}} \right)^{-5}.
\end{equation} 
Note that the IMF is independent of the mass of the pregalactic cloud 
(see eq. \ref{eq:uimf}).
Also the transition mass $m_{\rm tr} \simeq 40M_{\sun}$ does not depend
on the cloud mass because this comes from the discontinuity in the 
$m_{\rm frag}-W$ relation (Figure 2), which is 
independent of the cloud mass.

A bi-modal IMF for metal-free stars has also been proposed also by Nakamura \&
Umemura (2001), as mentioned in \S 1. It is interesing to note here the
differences between our IMF and theirs. They pointed out that the IMF becomes
bi-modal, with peaks around 1 and 100 $M_{\sun}$, if dense $(n_{\rm
H}>10^{6}{\rm cm^{-3}}$) filamentary clumps formed in the course of collapse. It
is, however, still uncertain that such dense filaments are formed in reality. On
the other hand, we consider here only the low-density filaments that fragment
initially during the loitering phase. Even so, the sub-solar mass peak appears,
owing to photodissociation feedback from pre-existing stars. Recall also that
the production mechanism of the low-mass stars is different in the two
treatments. Nakamura and Umemura (2001) considered fragmentation of dense
molecular-cooling clouds around $n_{\rm H} \simeq 10^{12} {\rm cm^{-3}}$, while
we considered fragmentation of atomic-cooling clouds around $n_{\rm H} \simeq
10^{16} {\rm cm^{-3}}$. 
 
\section{Metal Enrichment by First-Generation Stars}

The chemical enrichment by the first-generation stars has particular importance
for the subsequent chemical evolution of protogalaxies. Also, the masses of those
stars can be constrained by observing the abundance ratios of elements in
second-generation stars. In \S3, the IMF for metal-free stars was predicted to
be bi-modal with peaks around $0.3M_{\sun}$ and $40M_{\sun}$. In this
section, we discuss the chemical yield of the first-generation stars under the
predicted IMF.
 
Since low mass stars do not eject metals (until after completing their long
lifetimes), we do not need to know the detailed shape of the low-mass IMF. The
mass ratio of low-mass stars, $f_{\rm low}$, given by equation (\ref{eq:flow}),
suffices for our purpose. For simplicity, we approximate the high-mass portion
of IMF as  

\begin{equation}
\xi_{\rm high} \propto m_{\ast}^{-5}
~~~~~(m_{\rm tr}< m_{\ast} < m_{\rm first})        
\end{equation}

\noindent in this section.  We define the mean mass fraction of ejected 
element $X$ by supernovae  

\begin{equation}
\epsilon_{X,{\rm high}}=\int dm_{\ast} m_{X} \xi_{\rm high}
/\int dm_{\ast} m_{\ast} \xi_{\rm high},
\end{equation}

\noindent and the mean remnant mass fraction  

\begin{equation}
\alpha_{\rm high}=\int dm_{\ast} m_{\rm rem} \xi_{\rm high}
/\int dm_{\ast} m_{\ast} \xi_{\rm high}.
\end{equation}

\noindent Using these quantities, the yield of element X by the first-generation stars is expressed as

\begin{equation}
p_{X}=\frac{\epsilon_{X,{\rm high}} (1-f_{\rm low})} {f_{\rm low}+(1-f_{\rm
low}) \alpha_{\rm high}}.
\end{equation}
 
Considering the uncertainty of the transition mass, $m_{\rm tr}$, we calculate
the yield for two cases of $m_{\rm tr}=20$ and $40M_{\sun}$, respectively. We
fix $m_{\rm first}=2000M_{\sun}$ since the uncertainty of this causes little
difference owing to the steepness of IMF. The values of $\epsilon_{X,{\rm
high}}$ and $\alpha_{\rm high}$ are presented in Tables 1 and 2 for $m_{\rm
tr}=20$ and $40M_{\sun}$, respectively. We use the Population III supernova (SN)
yields by Umeda \& Nomoto (2002). Metals are ejected by stars in the mass range
$13-30M_{\sun}$ by Type II SNe, and $150-270M_{\sun}$ by pair-instability SNe.
Metal-free stars whose mass is between these ranges, as well as those more
massive than $270M_{\sun}$, collapse into black holes with little metal
ejection. Since the fraction of gas returning to the interstellar medium is
quite uncertain, we assume here that all of the stellar material is eventually
locked into black holes in this mass range.

A comparison of predicted elemental abundance ratios with the observed most
iron-deficient star HE~0107-5240 (Christlieb et al. 2002) is shown in Table 3.
When the cutoff mass, $m_{\rm tr}$, is higher than $30M_{\sun}$, metal
enrichment proceeds only via pair-instability SNe, and the abundance ratios
become identical to this single source. Christlieb et al. (2002) claimed that a
Type II SN of a $20-25M_{\sun}$ metal-free star reproduces the observed
abundance pattern heavier than Mg, while lighter elements could be accounted for
by self-enrichment or mass transfer from a more massive AGB companion star. On
the other hand, Schneider et al. (2003) pointed out that the abundance pattern
from a pair-instability SN of a $200-220M_{\sun}$ star also agrees with the
limited number of elemental abundances presently available for this star. So
far, the enrichment either by Type II or by pair-instability SNe seems to be
consistent with the observed abundance ratios. As a result, the abundance
pattern of HE~0107-5240 heavier than Mg is well-reproduced in all cases shown in
Table 3 within 0.2-0.3 dex, although in the case of $m_{\rm tr}>30M_{\sun}$ the
amount of Mg relative to Fe might be too small.

Recently, Umeda \& Nomoto (2003) proposed the ``mixing and fallback'' scenario:
the processed material is assumed to be mixed uniformly in the region from
$1.8M_{\sun}$ to $6.0M_{\sun}$, and almost all material in this layer falls back
to the central remnant; only a small fraction is ejected from this region.
Although this beautifully explains the abundance ratios of HE~0107-5240,
including the large C and O abundances relative to Fe, this theory contains
unknown free parameters, namely the extent of mixing and the mass-cut.

Note, however, that we do not take it as yet established that HE~0107-5240 is a
second-generation star formed from metal-enriched material. We stress here that
since sub-solar mass stars can form in a metal-free environment, HE~0107-5240
might be a first-generation star metal-polluted due to subsequent accretion
(Yoshii 1981; Shigeyama et al. 2003). The large [C/Fe](=+4.0) ratio of
HE~0107-5240, as well as the large [O/Fe] (=+2.4) ratio 
(private communication; M. S. Bessell), can be explained by the
self-enrichment in the AGB phase (private communication M. Y. Fujimoto;
Fujimoto, Ikeda, \& Iben 2000). The accretion scenario demonstrates that the gap
in the metallicity distribution function of observed low-metallicity stars,
between [Fe/H]=$-$5.3 and the current lower limit [Fe/H]= --3.5 to --4.0 of extreme
metal-poor stars could be real, as shown in Figure 2 of Shigeyama st al. (2003).

On the other hand, if the explosion of massive metal-free stars was responsible
for the birth of the [Fe/H]=--5.3 star as a second-generation star (Schneider et
al. 2003; Umeda \& Nomoto 2003; Bonifacio, Limongi, \& Chieffi 2003), there must
exist many more second-generation stars having a wide variety of metal
abundances below [Fe/H]=--3.5. Therefore, this massive Population III scenario
prefers the existence of a metallicity distribution without any appreciable gap
below [Fe/H]=--3.5, and future large samples of extremely metal-poor
stars will clearly discriminate between the competing scenarios proposed thus far.

\section{Summary and Discussion}

We have studied the initial mass function (IMF) of the first-generation stars
under the assumption that the fragmentation of prestellar clumps occurs around
the ``loitering phase,'' i.e., when the temperature stops decreasing. The
fragmentation mass scale decreases as the far ultraviolet radiation field
becomes intense. If the very first stars are very massive and create intense
radiation fields inside their parent clouds, stars formed later are smaller than
those formed earlier. The stellar initial mass function is established by this
mechanism. Low-mass star formation occurs only in clouds whose length scale $l$ is
less than the threshold length $l_{\rm th} \simeq 1.6 {\rm kpc}(M_{\rm
cl}/10^{8}M_{\sun})^{3/7}$, where $M_{\rm cl}$ is the initial gas mass of the
cloud. In such clouds, the IMF becomes bi-modal: the high-mass peak is about
$40M_{\sun}$ and the low-mass one locates at about $0.3M_{\sun}$. 
Numerous low-mass metal-free stars are formed, greatly exceeding 
the number of massive meta-free stars.
The high-mass portion of the IMF is found to be 
$\xi_{\rm high} = 3 \times 10^{-6} ({\rm stars}~M_{\sun}^{-1}~{\rm pc}^{-3})
(l/1{\rm kpc})^{-1} 
(m_{\ast}/50M_{\sun})^{-5}$, which
is a very steep function of stellar mass $m_{\ast}$, with a power-law index of
$-5$, in comparison with the Salpeter IMF's index of -2.35. Thus, although the
very first stars are possibly very massive, $\sim 1000M_{\sun}$, the typical
mass scale for metal-free stars is smaller, and depends on the length scale of
the pregalactic clouds.

If all metal-free stars were indeed as massive as the very first stars, they
would evolve to black holes without metal ejection. The surrounding gas remains
metal-free, and the very massive star formation mode continues indefinitely. This
is the so-called {\it star formation conundrum} pointed out by Schneider et al.
(2002). Our scenario, where the UV radiation from the preceding stars allows 
production of low-mass stars from metal-free gas,  
might provide a solution to this conundrum.

Recently, Mackey, Bromm, \& Hernquist (2003) studied the star formation history
in the high-redshift universe, in particular the transition from Population III
to Population II star formation. They assumed that the same top-heavy IMF is
maintained not only in low-mass halos that cool only via molecular cooling, but
also in halos that are massive enough to cool via atomic hydrogen lines.
According to our results, however, the mass scale of stars that form in massive
halos is smaller than those that form in low-mass halos.

We now understand that low-mass star formation occurs in a primordial cloud only
if the radius of the cloud is below the threshold $l_{\rm th} \simeq 1.6{\rm
kpc}(M_{\rm cl}/10^{8}M_{\sun})^{3/7}$. We should ask whether this is the case
in realisty. The virial radius of a pregalactic cloud, $r_{\rm vir} \simeq 2{\rm
kpc} (M_{\rm cl}/10^{8}M_{\sun})^{1/3} (1+z_{\rm vir}/20)^{-1}$, that collapses
at redshift $z_{\rm vir}$ should be regarded as the maximum radius of the cloud,
since it cools and contracts after virialization to reach dynamical equilibrium
as a whole. Comparing $l_{\rm th}$ and $r_{\rm vir}$, we conclude that
pregalactic clouds with $l<l_{\rm th}$ are not rare objects. Therefore, {\it
sub-solar mass metal-free stars are generically formed in the framework of
stardard cosmology. If metal-free low-mass stars are not be found in the
Galactic halo, an exotic scenario for structure formation must be invoked.}

Our model for pregalactic clouds is obviously oversimplified, and should
therefore be improved upon before the above conclusion is taken for granted. In
the following, we summarize our assumptions and their limitations. More
elaborate works taking these points into account are desirable in future studies. 
 
\begin{itemize}
 
\item
We have used a one-zone model for pregalactic clouds. Inhomogeneity inside a
pregalactic cloud is not taken into account. 
In particular, we have assumed that the star formation timescale is 
given by the averaged free-fall time.
Of course, the evolution of pregalactic clouds and initial mass function 
of first-generation stars crucially depend on this assumption.
Although this assumption has not been fully justified, there are some 
observational suggestions that the global star formation rate (SFR) 
in galaxies is indeed determined by galactic scale conditions 
(e.g., Kennicutt 1998).

First, we consider the effect of the spatial fluctuations of SFR 
within the pregalactic cloud.
If the star formation is very active in a region in comparison 
with the rest of the cloud, 
this star-forming region itself can be regarded as an almost independent 
cloud.
Since the length scale of the star-forming region is smaller 
than that of the entire cloud, this allows easier build-up of local 
radiation field and easier low-mass star fomation. 

Next, we discuss the effects of different values of the global SFR. 
Possible changes in the averaged SFR do not affect
the high-mass IMF directly, since the equation (\ref{eq:aimf}) does not 
contain the SFR.  
On the other hand, the condition for low-mass star formation discussed 
in \S 3.2 will be altered.
If the star formation process is slower than that assumed here, 
the radiation field will increase more slowly.
This results in a smaller value of the threshold length $l_{\rm th}$ 
for low-mass star formation than in equation (\ref{eq:lth}).

\item
We have only considered the averaged radiation field 
and have neglected its local fluctuations.
In reality, the radiation field is
stronger in a region closer to a massive star than in outer regions. 
Therefore,
dispersion around the average fragmentation scale must exist at any time. 
This effect might affect the IMF in the pregalactic clouds.
 
\item 
We have derived the fragmentation mass scale from the condition at the local
temperature minimum, or ``loitering point''. Because of this simplified
treatment, the value of fragmetation mass scale should be regarded as precise
only in the order-of-magnitude sense. More accurate determintation of
fragmentation mass scale under UV radiation field is an interesting topic for
future studies (see e.g., Bromm \& Loeb 2003; Yoshida et al. 2003). Also, the
fragmentation epoch depends on the initial density and line-mass of filaments
(Nakamura \& Umemura 2001). The dense and large line-mass filements collapse
until very high density ($10^{12}{\rm cm^{-3}}$) without fragmenting. It is
important to specify these properties of filaments in the realistic cosmological
context.

\item
We have assumed that the stellar mass scale coincides with the 
fragmentation scale.
If the accretion rate drops suddenly for massive protostars, as suggested by
Abel et al. (2002), the stellar mass scale falls short of the fragmentation
scale. This depends crucially on the late-time behavior of the accretion rate,
which has not yet been fully explored. 

\item
We have assumed that the ionizing photons are trapped in the HII region formed 
around the massive stars, and do not contribute to the average radiation 
field in the protogalaxies. 
This assumption depends on the size and distribution of the HII regions. 
From comparison of the results by Susa \& Kitayama (2000) and Omukai (2001), 
we expect that the thermal evolution does not deviate so much from 
those we used in \S 2 even if the ionizing photons are included.
Therefore, even if this assumption is invalidated, the results will 
not be altered significantly. 

\item
We have neglected kinematical effects of the HII regions formed around massive
stars. If either the density around the star or the protostellar accretion rate is
sufficiently high, the HII region is confined to a small region around the
exciting star. If not, the HII region expands due to  high internal pressure, and the
entire pregalactic cloud might be disrupted by photoevaporation. If the
pregalactic clouds are disk-like, an HII region extends preferentially in the
vertical direction to the disk, and it ends up with a hole like a chimney
(Franco, Tenorio-Tagle, \& Bodenheimer 1990). Therefore, even if
photoevaporation by massive stars occurs, it may not be sufficiently violent to disrupt
the entire pregalactic cloud. However, more thorough studies are necessary to
draw this conclusion with confidence.
 
\item
We have assumed that formed stars start shining as main sequence stars
immediately. Stars are, however, formed by protostellar accretion, and the
accretion phase precedes the main-sequence phase. If the mass accretion rate is
$\dot{m}_{\rm acc}$, the duration of accretion phase for a star of mass
$m_{\ast}$ is $t_{\rm acc}=m_{\ast}/ \dot{m}_{\rm acc}$. In the case of
primordial star formation, the accretion rate is very high:
$10^{-2}-10^{-3}M_{\sun}/{\rm yr}$ (Stahler, Palla, \& Salpeter 1986; Omukai \&
Nishi 1998). Then, the accretion phase is short in comparison with the stellar
lifetime. However, for the most massive stars ($\la 1000M_{\sun}$), the
accretion phase lasts for a significant part of the expected stellar lifetime. For such high
accretion rate, the accretion flow becomes optically thick to H$^{-}$ bound-free
absorption. The effective temperature, $T_{\rm eff}$, of accreting protostars is
about 6000K, while at the main-sequence stage it becomes $T_{\rm eff} \simeq 10^{5}K$ (Stahler
et al. 1986; Omukai \& Palla 2001, 2003). Thus, if the accretion phase is
prolonged, the emitted FUV radiation decreases significantly. 
 
\item
We have assumed that the star formation timescale is similar to the free-fall
time of the cloud. In other words, we considered only the spontaneous mode of
star formation. Triggered star formation by stellar winds or SNe (Elmegreen \&
Lada 1977) has not been included in our model, since both stellar winds or
pulsational mass losses are expected to be rather weak for metal-free stars
(Kudritzki 2000; Baraffe, Heger, \& Woosley 2001). 
 
\end{itemize}

\acknowledgements 
We thank Tim Beers for careful reading of the manuscript and 
the referee, Benedetta Ciardi, for comments that improved the presentation 
of the paper.
This work is supported in part by Research Fellowship of the Japan 
Society for the Promotion of Science for Young Scientists (6819; KO),
and the Grant-in-Aid for Scientific Research (1520401) and 
Center-of-Excellence Research (07CE2002) from Ministry of Education, 
Culture, Sports, Science, and Technology (YY).
\appendix
\newpage

\section{Thermal Instability in Collapsing Metal-Free Clumps}

In this appendix, we study the thermal-chemical instability using a linear
perturbation theory, following Sabano \& Yoshii (1977) 
and Yoshii \& Sabano (1979). The energy equation
(\ref{eq:energy}) and the equation of H$_2$ formation/dissociation 
(\ref{eq:H2form}) are perturbed under the isobaric condition $\delta P=0$. 
The energy equation is

\begin{equation}
\frac{dU}{dt}-\frac{P}{\rho}\frac{d{\rm ln} \rho}{dt}=-{\cal L}
\label{eq:energy}
\end{equation}

\noindent where $U$ is the internal enegy per unit mass, including the thermal and 
chemical binding energy
\begin{equation}
U=\frac{3}{2}\frac{k_{\rm B}T}{\mu_{\rm a} m_{\rm H}}
-\frac{1}{2}\frac{\chi f}{m_{\rm H}},
\label{eq:u}
\end{equation}
and ${\cal L}={\cal L}_{\rm H_2}$ is the cooling rate by H$_2$ line emission.
In equation (\ref{eq:u}), $f$ is the molecular fraction 
\footnote{$f=1$ corresponds to fully molecular.}, and $\chi$ is the chemical binding 
energy of H$_2$.
Note that, unlike in \S 2, the cooling rate ${\cal L}$ does not include 
chemical cooling/heating. 
Instead, the definition of the internal energy $U$ is modified to 
include the chemical binding energy.
The equation of state for an ideal gas is used: 
\begin{equation}
P=\frac{\rho k_{\rm B} T}{\mu m_{\rm H}}.
\end{equation}
In the equations above, 
\begin{equation}
\mu^{-1}=1-\frac{f}{2}, 
\end{equation}
and 
\begin{equation}
\mu_{\rm a}^{-1}=1-\frac{3 \gamma_{\rm H_2} -4}{\gamma_{\rm H_2} -1} 
\frac{f}{3}=1-\frac{f}{6},
\end{equation}
where the adiabatic index for molecular hydrogen, $\gamma_{\rm H_2}=7/5$,
has been used, and the electron fraction, $x_{e}$, has been assumed to be small,
$x_{e} \ll 1$.
 
The equation for H$_{2}$ formation/dissociation is
\begin{equation}
\frac{df}{dt}={\cal F},
\label{eq:H2form} 
\end{equation}
where ${\cal F}$ is the net formation rate for H$_2$.
 
By introducing new variables
\begin{equation}
x=\mu^{-1}=1-\frac{f}{2}, 
\end{equation}
\begin{equation}
{\cal P}=\mu_{\rm a} [\frac{2m_{\rm H} {\cal L}}
{3k_{\rm B}T}-\frac{1}{3}(\frac{1}{2}+\frac{\chi}{k_{\rm B}T}){\cal F} ], 
\end{equation}
and 
\begin{equation}
{\cal Q}=\frac{{\cal F}}{2-f},
\end{equation}
the equations (\ref{eq:energy}) and (\ref{eq:H2form}) can be written as
\begin{equation}
\frac{d{\rm ln}T}{dt}-\frac{2x}{x+2} \frac{d{\rm ln}\rho}{dt} + {\cal P} =0 
\label{eq:energy2}
\end{equation}
and 
\begin{equation}
\frac{d{\rm ln}x}{dt}+ {\cal Q}=0.
\label{eq:H2form2}
\end{equation}
 
For perturbed quantities, we use the following notation:
\begin{equation}
\delta_{T}=\frac{\delta T}{T},~~ 
\delta_{\rho}=\frac{\delta \rho}{\rho},~~
\delta_{x}=\frac{\delta x}{x}.
\end{equation}
 
The linear analysis is carried out under the isobaric condition 
$\delta P=0$:
\begin{equation}
\delta_{T}+\delta_{\rho}+\delta_{x}=0.
\label{eq:isobaric}
\end{equation}
 
Perturbation of equation (\ref{eq:energy2}) reads:
\begin{equation}
\frac{d \delta_{T}}{dt}-\left(\frac{2x}{x+2}\right)\frac{d \delta_{\rho}}{dt}
+T{\cal P}_{T} \delta_{T}+ \rho{\cal P}_{\rho} \delta_{\rho}
+\left(x{\cal P}_{x}-{\cal R}\right) \delta_{x}=0,
\label{eq:pertenergy}
\end{equation}
where 
\begin{equation}
{\cal R}=\frac{4x}{(x+2)^2}t_{\rm dyn}^{-1}.
\end{equation}
We define the dynamical timescale as
\begin{equation}
t_{\rm dyn}=\left(\frac{d {\rm ln} \rho}{dt} \right)_{\rm bg}^{-1}
\end{equation}
for the collapsing background of the perturbation.
 
Perturbation of H$_2$ formation/dissociation equation reads:
\begin{equation}
\frac{d \delta_{x}}{dt}+ T{\cal Q}_{T} \delta_{T}
+ \rho{\cal Q}_{\rho} \delta_{\rho}+ x{\cal Q}_{x} \delta_{x} =0.
\label{eq:perth2}
\end{equation}
 
We assume that $\delta_{T}, \delta_{\rho}$, and 
$\delta_{x} \propto {\rm exp} (\omega t)$.
Seeking non-trivial solutions of equations (\ref{eq:isobaric}), 
(\ref{eq:pertenergy}) and (\ref{eq:perth2})
for $\delta_{T}, \delta_{\rho}$ and $\delta_{x}$, 
we obtain the following equation for the growth rate of instabiliby, $\omega$:
\begin{equation}
a_{0} \omega^{2} + a_{1} \omega + a_{2} = 0.
\label{eq:omega}
\end{equation}
The coefficients are
\begin{equation}
a_{0}=1+\frac{3}{2}\frac{\mu}{\mu_{\rm a}},
\end{equation}
\begin{eqnarray}
a_{1}&=&\left( \frac{x+2}{2x} \right)
\left[ (T {\cal P}_{T}-\rho {\cal P}_{\rho})
+(x {\cal Q}_{x}- \rho {\cal Q}_{\rho}) \right]
-(T {\cal Q}_{T}-x {\cal Q}_{x}) \\
&=&\frac{\mu m_{\rm H}}{k_{\rm B}T} 
\left(T {\cal L}_{T}- \rho {\cal L}_{\rho}-{\cal L} \right) 
-\frac{\mu \chi}{2 k_{\rm B} T} 
\left(T{\cal F}_{T}-\rho{\cal F}_{\rho}-{\cal F} \right) \\
&~&-\left(1+\frac{3}{2}\frac{\mu}{\mu_{\rm a}} \right) 
\left({\cal F}_{f} +\frac{\mu}{2} \rho {\cal F}_{\rho}
+\frac{{\cal F}}{2-f}\right)
-\frac{3 \mu}{4}\left(T {\cal F}_{T}- \rho {\cal F}_{\rho} \right),
\end{eqnarray}
and
\begin{eqnarray}
a_{2}&=&\left( \frac{x+2}{2x} \right)
\left[ (T {\cal P}_{T}-\rho {\cal P}_{\rho})  
(x {\cal Q}_{x}-\rho {\cal Q}_{\rho})
-(T {\cal Q}_{T}-\rho {\cal Q}_{\rho})  
\left(x {\cal P}_{x}-\rho {\cal P}_{\rho}
- {\cal R} \right) \right] \\
&=&-\frac{\mu m_{\rm H}}{k_{\rm B}T} 
\left(T {\cal L}_{T}- \rho {\cal L}_{\rho} -{\cal L}\right)
\left({\cal F}_{f}+\frac{\mu}{2} \rho {\cal F}_{\rho} 
+\frac{{\cal F}}{2-f}\right) \\
&~&+\frac{\mu m_{\rm H}}{k_{\rm B}T}
\left({\cal L}_{f}+\frac{\mu}{2} \rho {\cal L}_{\rho}
+\frac{\mu_{\rm a}}{6} {\cal L} \right)
\left(T {\cal F}_{T}- \rho {\cal F}_{\rho}\right) \\
&~&+\frac{\mu}{2}{\cal F}\left[ \frac{\mu \mu_{\rm a}}{3} 
\left( \frac{1}{2}+ \frac{\chi}{k_{\rm B}T}\right)
\left(T {\cal F}_{T}- \rho {\cal F}_{\rho}\right) 
-\frac{\chi}{k_{\rm B}T} \left({\cal F}_{f}
+\frac{\mu}{2} \rho {\cal F}_{\rho} +\frac{{\cal F}}{2-f} \right) \right]\\
&~&+\frac{\mu \mu_{\rm a}}{3 t_{\rm dyn}} 
\left(T {\cal F}_{T}- \rho {\cal F}_{\rho}\right) .
\end{eqnarray}

In the bottom panel of Figure 6, we plot the e-folding growth time 
of the instability $t_{\rm e}=\omega^{-1}$ for a collapsing primordial clump.
The collapse rate is assumed to be free-fall, 
$t_{\rm dyn}=t_{\rm ff}=\sqrt{3 \pi/ 32G \rho}$.
The H$_2$ line cooling rate is computed as in Omukai (2001), where the 
escape probability method is used for the cooling by optically 
thick lines.  
Also shown are the background temperature ({\it top}) and 
molecular fraction ({\it middle}) for the perturbation.  
We see from this figure that fluctuations grow by thermal 
instability, in other words, the thermal instablity timescale, $t_{\rm e}$, 
falls below the collapse timescale for the clump, $t_{\rm ff}$, 
during the active phase of three-body H$_2$ formation process.
However, the ratio of the growth rate relative to the to collapse rate,
$t_{\rm ff}/t_{\rm e}$, remains below 2.
Since the scale for thermal instability is $\lambda_{\rm th} 
\sim c_{\rm s}t_{\rm e}$ while the core size is approximately
the Jeans length, $\lambda_{\rm J} \sim c_{\rm s}t_{\rm ff}$, 
this means that the growing scale is larger than a half of the core size.
Therefore, although the thermal instability does occur, it may only 
accelerate the collapse of the central region rather than driving 
further fragmentation. 

Abel et al. (2000) investigated the collapse of a prestellar core 
by 3-D hydrodynamical simulation.
In their calculation, the sub-fragmentation during the 
three-body H$_2$ formation has not been observed. 
They interpreted this as a result of hypothetical turbulent mixing.
Our argument here may give an alternative explanation 
of their result.


\newpage
\bigskip
\centerline{\bf Figures}
 
\plotone{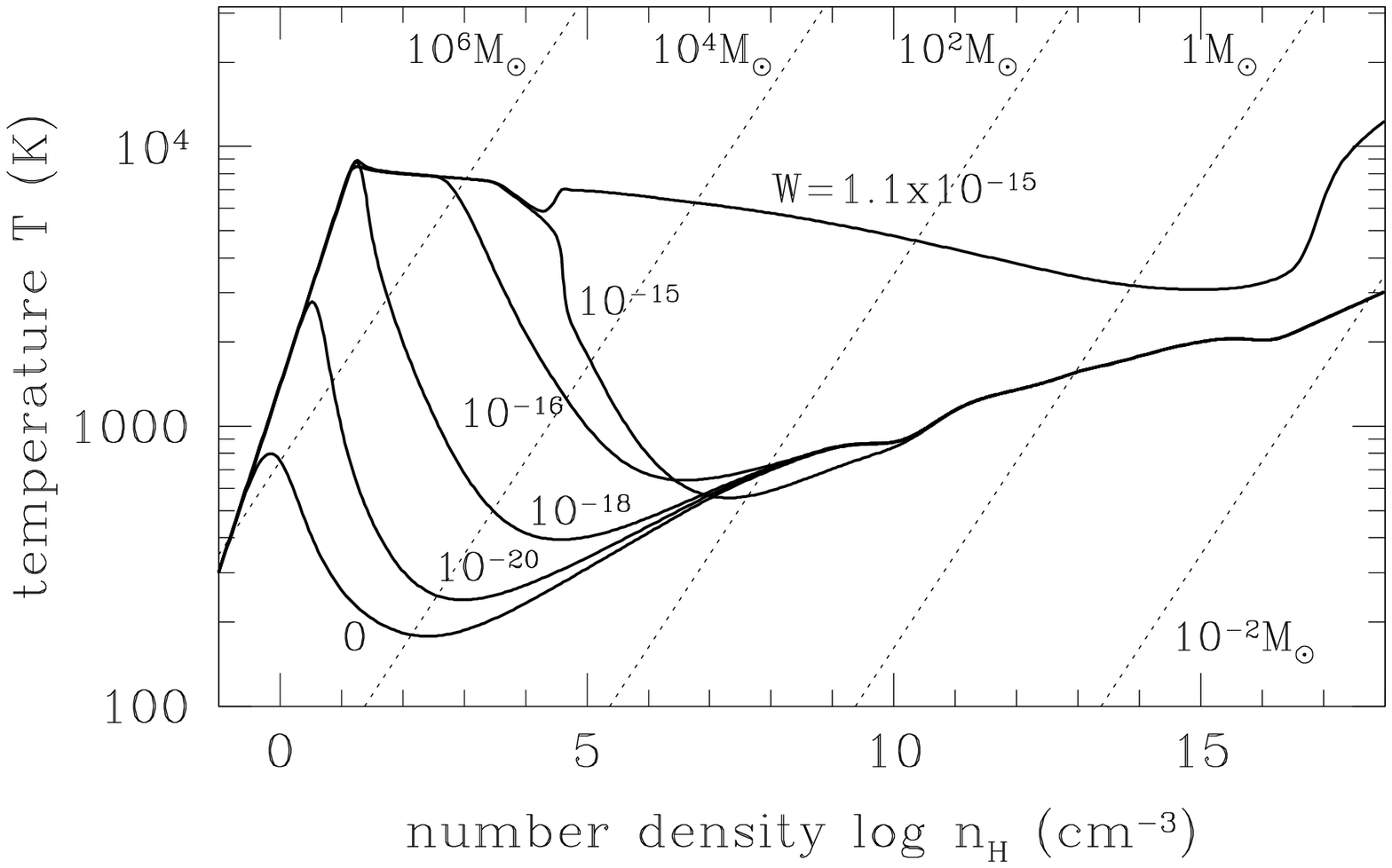}
\figcaption[f1.eps]{The temperature evolution for contracting metal-free 
prestellar clumps as a function of FUV radiation density.
FUV radiation density is parameterized by the dilution factor 
$W=u_{\rm rad}/aT_{\rm rad}^{4}$, where $T_{\rm rad}=10^5$K.
The values of $W$ are denoted in the figure.
The collapse proceeds by H$_2$ cooling for FUV radiation below 
the threshold $W_{\rm noH_{2}} \simeq 1.1 \times 10^{-15}$.
Whereas for $W>W_{\rm noH_{2}}$, H$_2$ formation is prohibited 
by photodissocitation, and the clump collapses by atomic cooling.
The dashed lines indicate the constant Jeans mass.
\label{fig:f1}}
 
\plotone{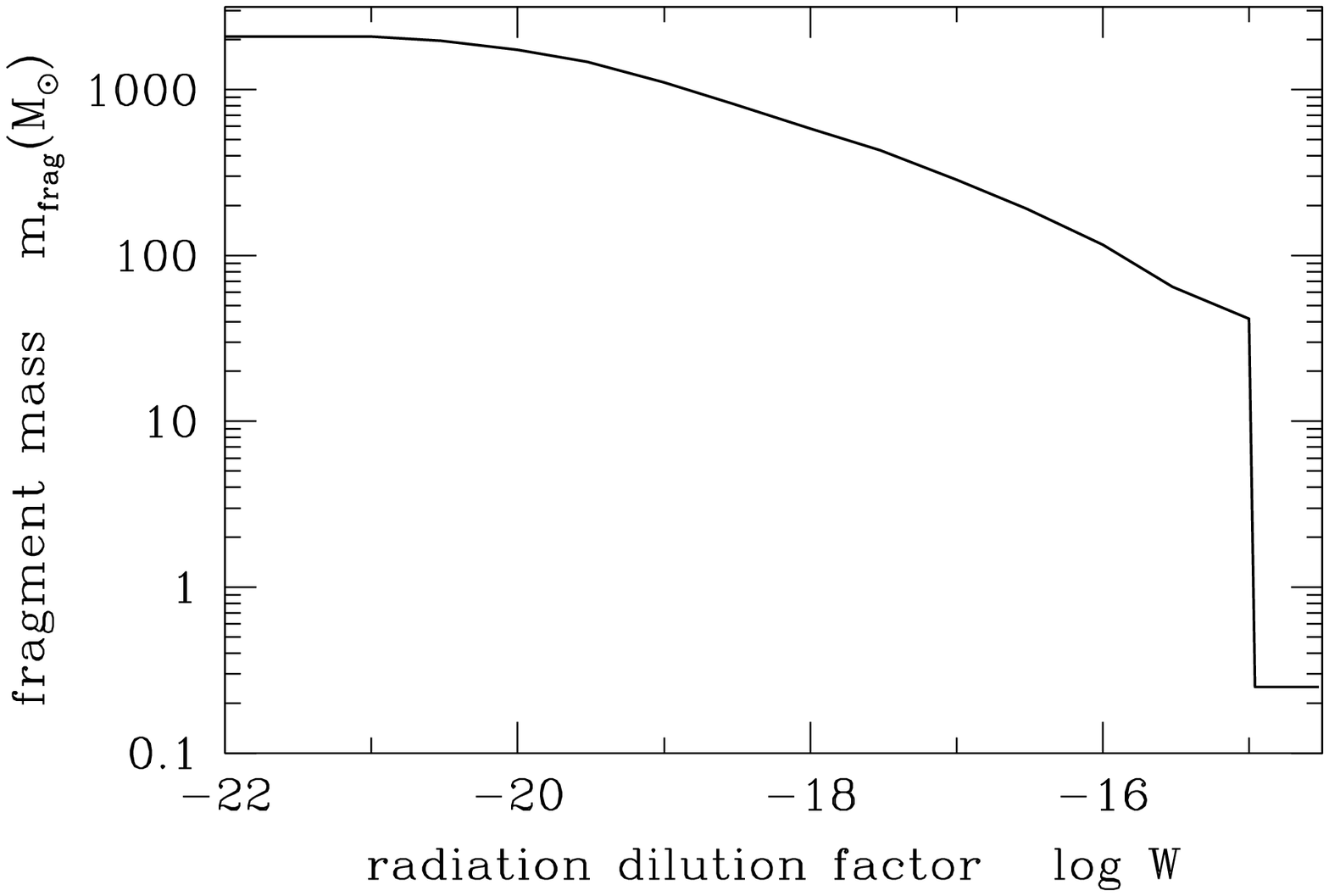}
\figcaption[f2.eps]{The fragmentation mass scale as a function of
the dilution factor $W$, which parametrizes the FUV radiation density.
\label{fig:f2}}
 
\plotone{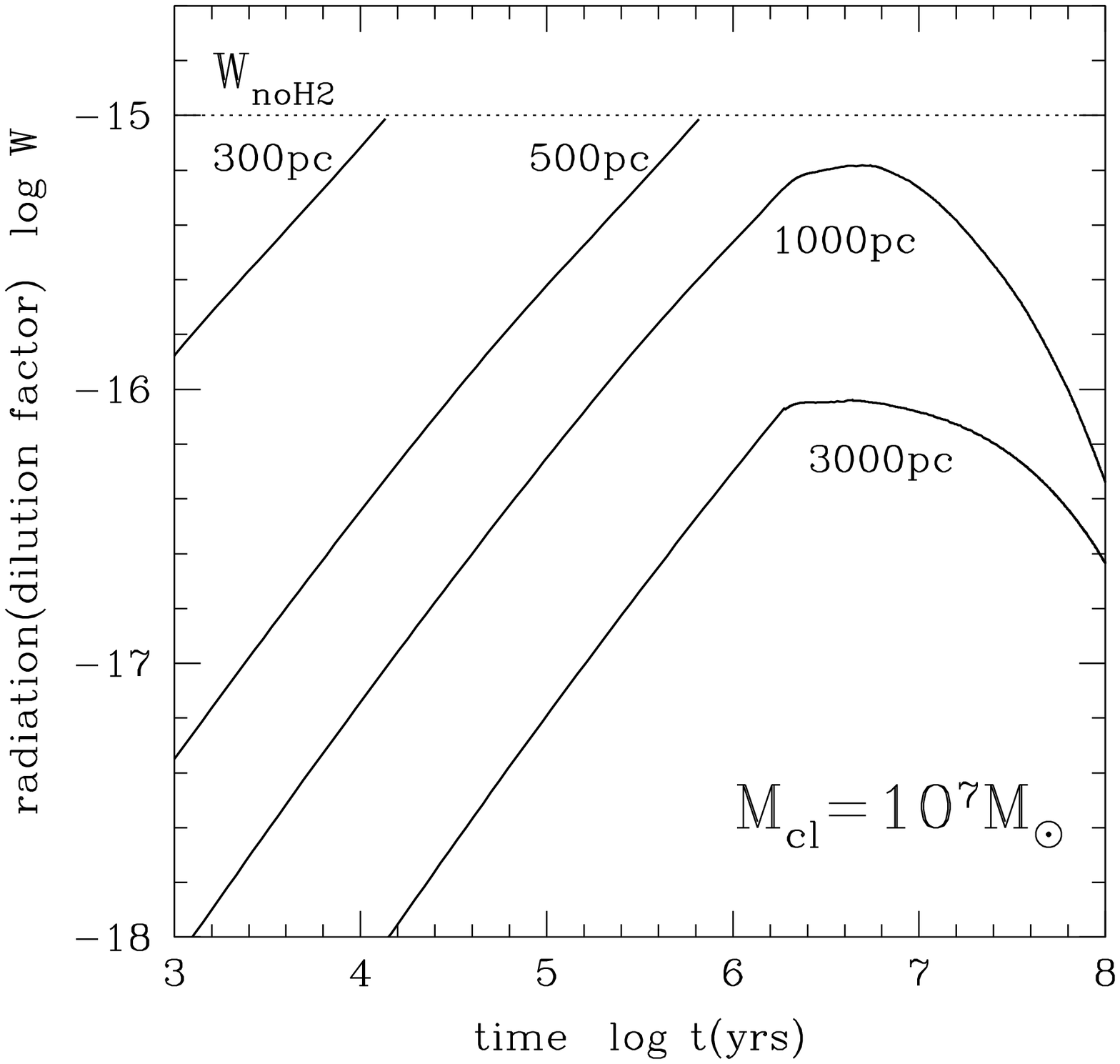}
\figcaption[f3.eps]{The evolution of FUV ratiation density as a
function of the cloud length scale. The initial gas mass of the clouds is
$M_{\rm cl}=10^{7}M_{\sun}$.
\label{fig:f3}}
 
\plotone{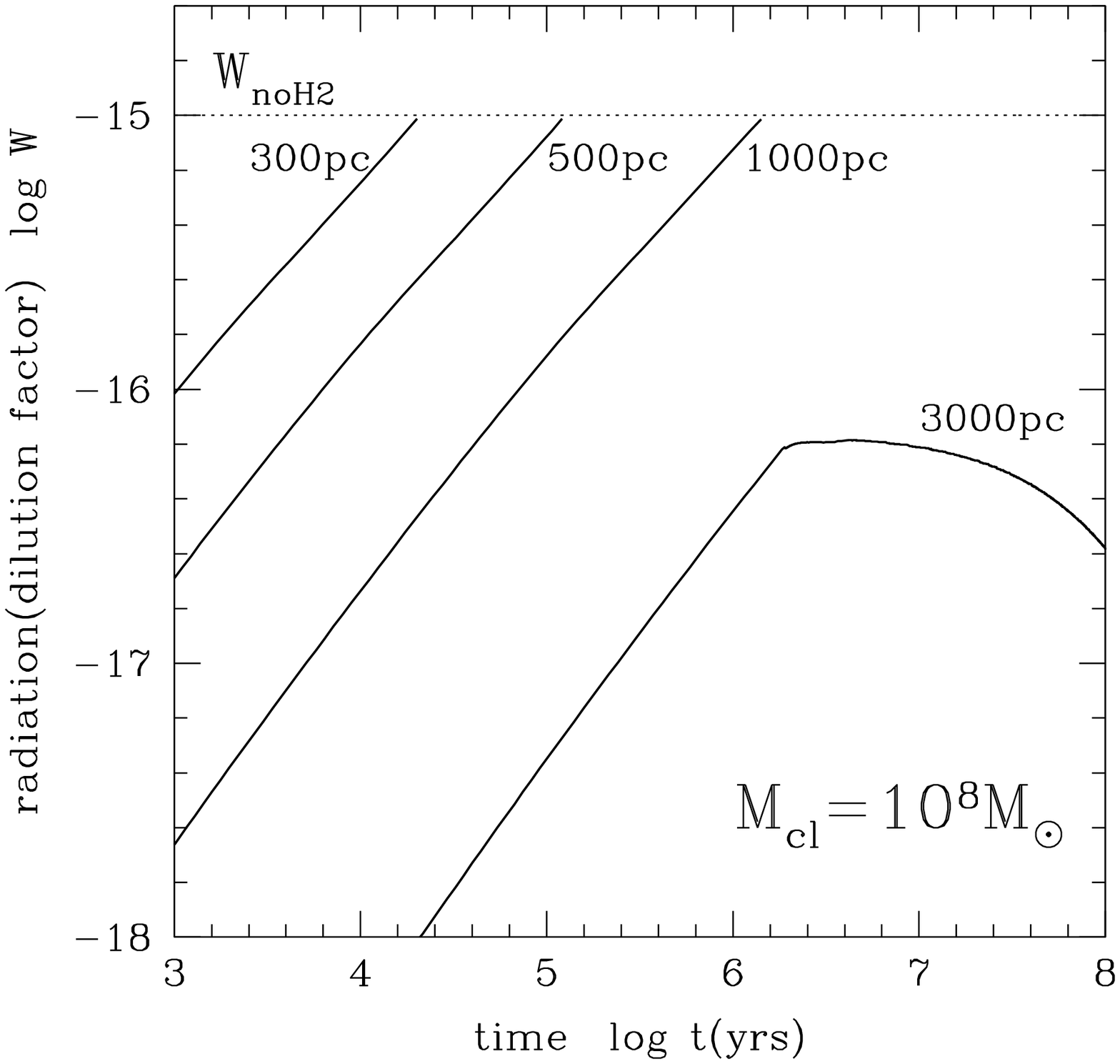}
\figcaption[f4.eps]{The same as Fig.3, but for the initial gas mass 
$M_{\rm cl}=10^{8}M_{\sun}$.
\label{fig:f4}}
 
\plotone{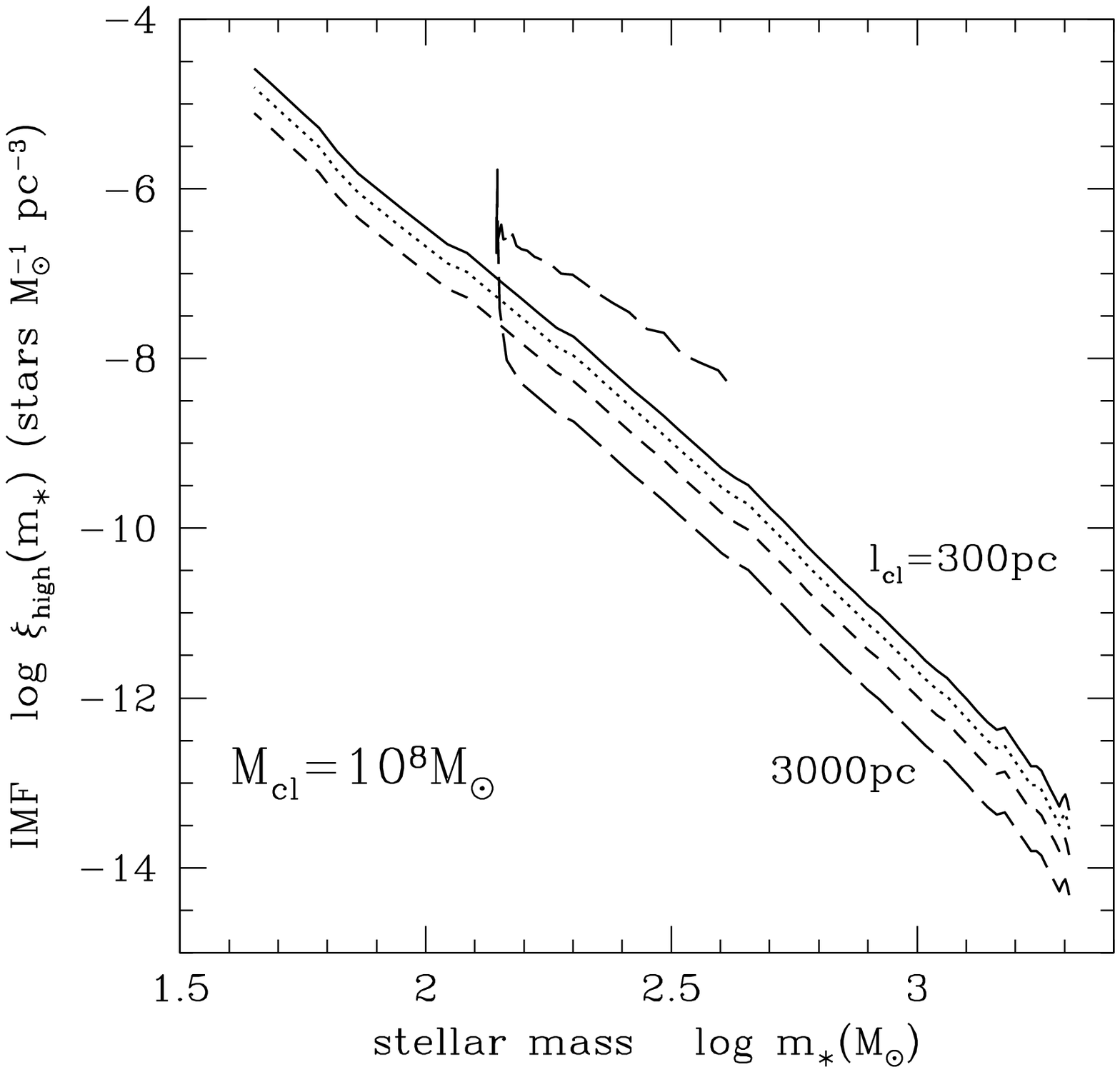}
\figcaption[f5.eps]{The initial mass function of metal-free stars in clouds 
of $M_{\rm cl}=10^{8}M_{\sun}$. 
 Shown are the results for the cloud length scale of 300pc (solid line), 
500pc (dotted line), 1000pc (short-dashed line), and 3000pc (long-dashed line). 
\label{fig:f5}}
 
\plotone{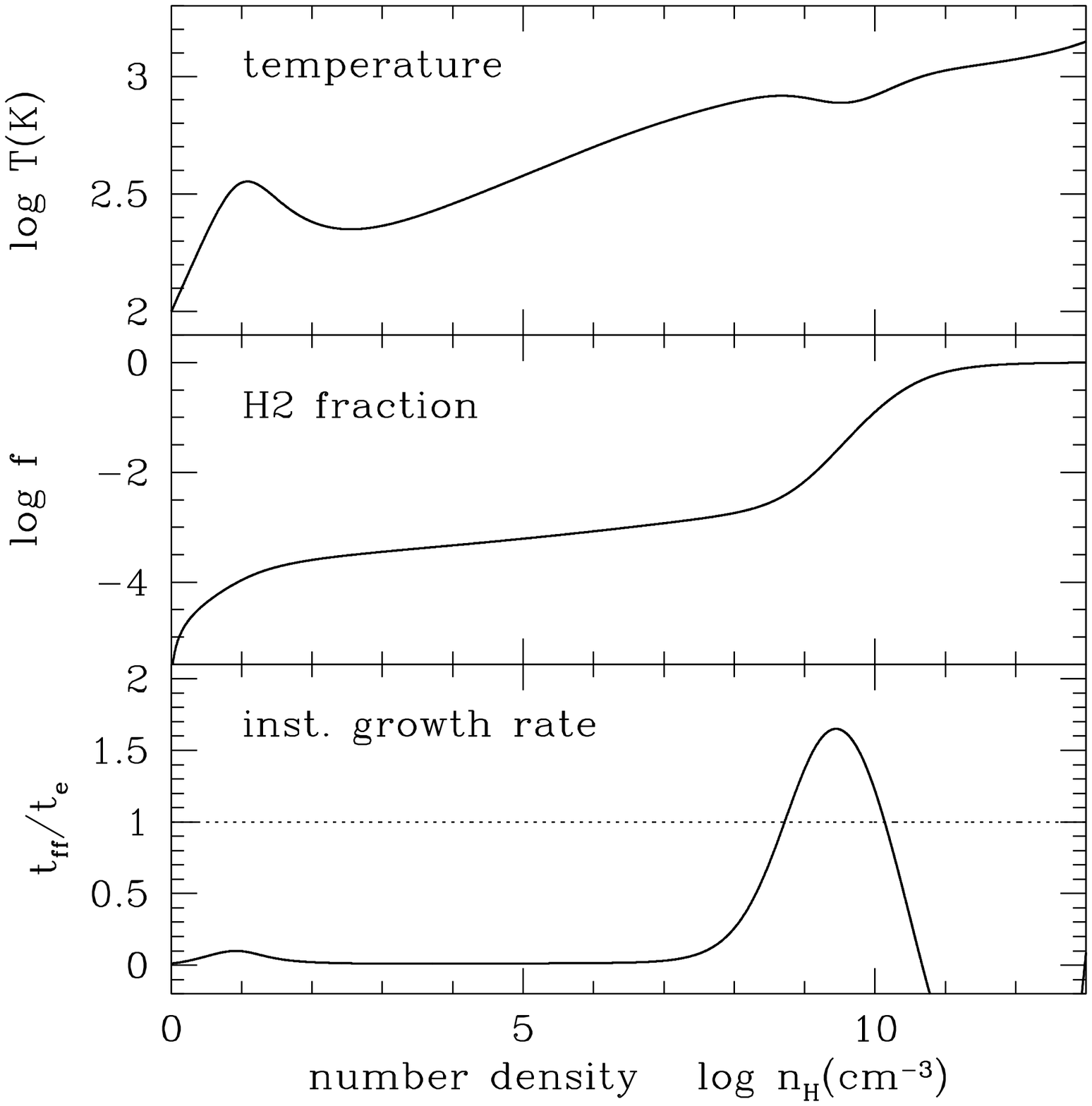}
\figcaption[f6.eps]{The growth rate of thermal instability 
($t_{\rm e}^{-1}=\omega$) relative to the free-fall rate along 
the evolutionary path of the collapsing clump.  Also shown are the gas
temperature and H$_2$ fraction evolutions used as
the unperturbed state of such clump.
\label{fig:f6}}

\begin{table}
\begin{center}
\renewcommand{\arraystretch}{1.2}
\begin{tabular}{rcrcrcrcrc} \hline\hline
 
\hline
 
     p   &    3.11E-01&  d       &    3.91E-08& $^3$He   &    9.28E-07&
$^4$He   &    2.20E-01& $^6$Li   &    1.05E-12\\
$^7$Li   &    9.41E-11& $^9$Be   &    2.62E-20& $^{10}$B &    2.21E-11&
$^{11}$B &    9.48E-11& $^{12}$C &    1.24E-02\\
$^{13}$C &    1.32E-09& $^{14}$N &    9.07E-06& $^{15}$N &    3.42E-08&
$^{16}$O &    6.53E-02& $^{17}$O &    9.72E-09\\
$^{18}$O &    1.78E-07& $^{19}$F &    9.00E-09& $^{20}$Ne&    5.28E-03&
$^{21}$Ne&    6.53E-07& $^{22}$Ne&    1.20E-06\\
$^{23}$Na&    1.72E-05& $^{24}$Mg&    3.27E-03& $^{25}$Mg&    2.39E-06&
$^{26}$Mg&    1.51E-06& $^{26}$Al&    5.16E-08\\
$^{27}$Al&    3.43E-05& $^{28}$Si&    4.44E-03& $^{29}$Si&    1.38E-05&
$^{30}$Si&    3.77E-06& $^{31}$P &    2.23E-06\\
$^{32}$S &    1.87E-03& $^{33}$S &    4.59E-06& $^{34}$S &    1.54E-06&
$^{36}$S &    1.34E-11& $^{35}$Cl&    3.66E-07\\
$^{37}$Cl&    3.50E-07& $^{36}$Ar&    3.09E-04& $^{38}$Ar&    3.77E-07&
$^{40}$Ar&    1.48E-13& $^{39}$K &    1.15E-07\\
$^{40}$K &    1.41E-11& $^{41}$K &    3.85E-08& $^{40}$Ca&    2.74E-04&
$^{42}$Ca&    6.58E-09& $^{43}$Ca&    1.89E-09\\
$^{44}$Ca&    2.00E-06& $^{46}$Ca&    3.97E-14& $^{48}$Ca&    8.60E-15&
$^{45}$Sc&    1.75E-09& $^{46}$Ti&    6.86E-09\\
$^{47}$Ti&    3.03E-08& $^{48}$Ti&    4.98E-06& $^{49}$Ti&    8.81E-08&
$^{50}$Ti&    1.25E-14& $^{50}$V &    5.29E-14\\
$^{51}$V &    6.63E-08& $^{50}$Cr&    4.60E-08& $^{52}$Cr&    4.70E-05&
$^{53}$Cr&    1.82E-06& $^{54}$Cr&    1.37E-12\\
$^{55}$Mn&    2.25E-06& $^{54}$Fe&    7.04E-06& $^{56}$Fe&    2.79E-03&
$^{57}$Fe&    4.77E-05& $^{58}$Fe&    5.61E-12\\
$^{59}$Co&    3.78E-07& $^{58}$Ni&    5.50E-06& $^{60}$Ni&    7.57E-05&
$^{61}$Ni&    2.48E-06& $^{62}$Ni&    1.54E-06\\
$^{64}$Ni&    5.73E-14& $^{63}$Cu&    3.58E-08& $^{65}$Cu&    3.53E-08&
$^{64}$Zn&    3.82E-06& $^{66}$Zn&    1.24E-07\\
$^{67}$Zn&    1.63E-10& $^{68}$Zn&    1.40E-09& $^{70}$Zn&    2.74E-14&
$^{69}$Ga&    2.01E-10& $^{71}$Ga&    9.98E-14\\
$^{70}$Ge&    4.51E-10& $^{72}$Ge&    2.11E-13& $^{73}$Ge&    1.05E-13&
$^{74}$Ge&    6.10E-14& &\\
 
\hline 
\end{tabular}
\end{center}
\caption{The mean ejected mass fraction of element X, $\epsilon_{\rm X,high}$, 
in the case of the transition mass $m_{\rm tr}=20M_{\sun}$.
The mean remnant mass is $\alpha_{\rm high}=0.379$}
\end{table}

\begin{table}
\begin{center}
\renewcommand{\arraystretch}{1.2}
\begin{tabular}{rcrcrcrcrc} \hline\hline
 
\hline
 
     p   &    3.58E-03&  d       &    9.53E-11& $^3$He   &    4.84E-09&
$^4$He   &    4.36E-03& $^6$Li   &    2.35E-20\\
$^7$Li   &    1.69E-13& $^9$Be   &    7.85E-20& $^{10}$B &    2.14E-19&
$^{11}$B &    6.49E-14& $^{12}$C &    3.21E-04\\
$^{13}$C &    7.82E-10& $^{14}$N &    6.68E-07& $^{15}$N &    2.09E-08&
$^{16}$O &    4.40E-03& $^{17}$O &    1.86E-08\\
$^{18}$O &    8.13E-08& $^{19}$F &    2.33E-11& $^{20}$Ne&    2.33E-04&
$^{21}$Ne&    3.21E-08& $^{22}$Ne&    6.18E-08\\
$^{23}$Na&    6.28E-07& $^{24}$Mg&    2.38E-04& $^{25}$Mg&    2.39E-07&
$^{26}$Mg&    7.67E-08& $^{26}$Al&    9.99E-09\\
$^{27}$Al&    2.19E-06& $^{28}$Si&    1.34E-03& $^{29}$Si&    1.94E-06&
$^{30}$Si&    2.19E-07& $^{31}$P &    1.76E-07\\
$^{32}$S &    7.43E-04& $^{33}$S &    8.74E-07& $^{34}$S &    1.80E-07&
$^{36}$S &    1.82E-12& $^{35}$Cl&    7.20E-08\\
$^{37}$Cl&    1.72E-07& $^{36}$Ar&    1.24E-04& $^{38}$Ar&    1.47E-07&
$^{40}$Ar&    9.46E-15& $^{39}$K &    4.24E-08\\
$^{40}$K &    1.38E-12& $^{41}$K &    2.65E-08& $^{40}$Ca&    1.20E-04&
$^{42}$Ca&    4.05E-09& $^{43}$Ca&    1.92E-09\\
$^{44}$Ca&    5.47E-08& $^{46}$Ca&    5.65E-14& $^{48}$Ca&    4.50E-17&
$^{45}$Sc&    2.35E-09& $^{46}$Ti&    2.44E-09\\
$^{47}$Ti&    1.79E-09& $^{48}$Ti&    5.20E-07& $^{49}$Ti&    1.75E-08&
$^{50}$Ti&    1.90E-16& $^{50}$V &    4.76E-15\\
$^{51}$V &    1.44E-08& $^{50}$Cr&    3.95E-08& $^{52}$Cr&    9.19E-06&
$^{53}$Cr&    4.73E-07& $^{54}$Cr&    4.12E-12\\
$^{55}$Mn&    1.58E-06& $^{54}$Fe&    1.55E-05& $^{56}$Fe&    3.31E-04&
$^{57}$Fe&    4.32E-06& $^{58}$Fe&    3.54E-11\\
$^{59}$Co&    1.39E-07& $^{58}$Ni&    1.32E-05& $^{60}$Ni&    1.23E-06&
$^{61}$Ni&    8.15E-09& $^{62}$Ni&    3.01E-08\\
$^{64}$Ni&    4.28E-16& $^{63}$Cu&    2.14E-10& $^{65}$Cu&    1.34E-11&
$^{64}$Zn&    1.42E-09& $^{66}$Zn&    2.41E-10\\
$^{67}$Zn&    5.82E-13& $^{68}$Zn&    1.92E-13& $^{70}$Zn&    3.24E-16&
$^{69}$Ga&    4.01E-14& $^{71}$Ga&    8.33E-16\\
$^{70}$Ge&    8.42E-13& $^{72}$Ge&    1.49E-15& $^{73}$Ge&    1.08E-15&
$^{74}$Ge&    3.55E-16& &\\
 
\hline 
\end{tabular}
\end{center}
\caption{The same as Table 1, but for the transition mass 
$m_{\rm tr}=40M_{\sun}$. 
The mean remnant mass fraction $\alpha_{\rm high}=0.984$}
\end{table}
 
\begin{table}
\begin{center}
\renewcommand{\arraystretch}{1.2}
\begin{tabular}{lcccccccccc} \hline\hline
[X/Fe]      &   Li  &   C  &   N  &  O  &  Na &  Mg  &  Ca &  Ti  &  Ni  &  Zn \\
\hline \hline
HE~0107-5240 & $<5.3$ &  4.0 &  2.3 &  2.4  & 0.8 &  0.2 & 0.4 & -0.4 & -0.4 
& $<2.7$ \\
\hline
$m_{\rm tr}=20M_{\sun}$ 
& -3.28 & -0.41 & -3.03 & -0.06 & -1.02 & -0.02 & 0.16 & -0.17 & -0.26 & 0.00\\
$m_{\rm tr}=25M_{\sun}$
& -4.43 & -0.49 & -3.21 & -0.05 & -1.12 & 0.00 & 0.18 & -0.18 & -0.26 & -0.08\\
$m_{\rm tr}>30M_{\sun}$
& -5.12 & -1.09 & -3.25 & -0.32 & -1.55 & -0.25 & 0.70& -0.23 & -0.12 & -2.46\\
\hline 
\end{tabular}
\end{center}
\caption{Comparison of abundance ratios [X/Fe] with HE~0107-5240}
\end{table}
 
\end{document}